\documentclass{amsart}

\usepackage{graphicx}
\usepackage[ruled,vlined]{algorithm2e}
\usepackage{amssymb} 

\newtheorem{theorem}{Theorem}[section]

\theoremstyle{definition}
\newtheorem{definition}[theorem]{Definition}

\theoremstyle{remark}

\numberwithin{equation}{section}

\usepackage{eqnarray,amsmath}

\newcommand{\R}{\mathbb{R}}

\newcommand{\E}{\mathbb{E}}
\renewcommand{\H}{\mathbb{H}}
\renewcommand{\S}{\mathbb{S}}
\newcommand{\T}{\mathbb{T}}

\newcommand{\eNorm}[1]{\lvert#1\rvert_{\E}}

\newcommand{\eDot}[2]{\langle#1,#2\rangle_{\mathbb{E}}}
\newcommand{\hDot}[2]{\langle#1,#2\rangle_{\mathbb{H}}}

\begin{document}

\title{Global Illumination of non-Euclidean spaces}

\author{Tiago Novello}
\address{VISGRAF Laboratory,Rio de Janeiro,Brazil}
\email{tiago.novello90@gmail.com}
\author{Vinicius da Silva}
\address{VISGRAF Laboratory,Rio de Janeiro,Brazil}
\email{dsilva.vinicius@gmail.com}
\author{Luiz Velho}
\address{VISGRAF Laboratory,Rio de Janeiro,Brazil}
\email{lvelho@impa.br}


\date{June 24, 2019}



\begin{abstract}
This paper presents a path tracer algorithm to compute the global illumination of non-Euclidean manifolds. We use the 3D torus as an example.  
\end{abstract}

\maketitle
\begin{figure}[hh]
	\centering
	\includegraphics[width=4.3in]{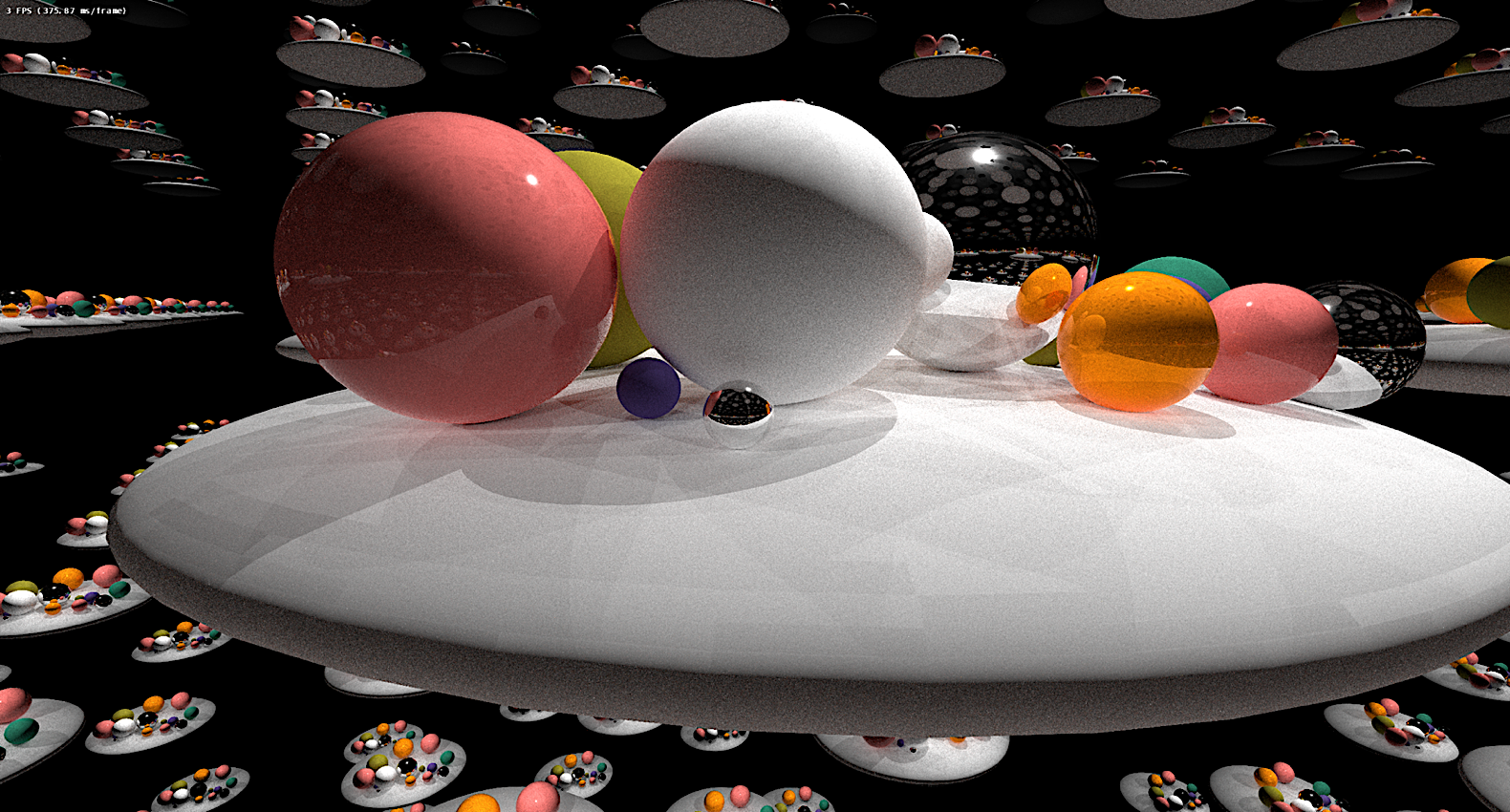}
	\label{fig:six_flat_manifolds}
\end{figure}


\section{Introduction} \label{sec:introduction}

\textit{Global illumination} is a collection of algorithms in computer graphics that are employed to mimic realist lighting in 3D scenes. The complexity of these algorithms depends on the \textit{indirect illumination}, which comes not directly from the light sources (\textit{direct illumination}) but the light of these sources reflected by other surfaces at the scene. Then computing \textit{reflections}, \textit{refractions}, and \textit{shadows} are important to compute the global illumination of a given scene. 

In \textit{photorealistic rendering} a set of techniques are used to create an image from a 3D scene that approximates from a photograph~\cite{pharr2016physically}, thus enabling the visualization of the global illumination of the underlying 3D scene. \textit{Path tracing} is an algorithm contained in this set. Informally, it consists of giving color for each point (\textit{eye}) and direction (\textit{pixel}) by launching a path of rays through the scene based on reflections, refractions, and shadows. 
To write a ray tracer it is common to simulate \textit{cameras}, \textit{ray--surface intersections}, \textit{light sources}, \textit{visibility}, \textit{surface material}, \textit{indirect light transport}, and \textit{ray propagation}.

The above concepts used in global illumination of Euclidean spaces can be extended to non-Euclidean spaces.
These abstract spaces were developed avoiding Euclid's fifth postulate, they do ``not'' exist in the real world. Therefore no real camera can take a picture in such spaces, which makes exploring photorealistic rendering very challenging. Attempts to visualize non-Euclidean geometries have produced beautiful and inspiring images with no counterpart in the real world; much of this is due to the non-trivial topology and geometry of such abstract spaces. 

\begin{figure}[hh]
	\centering
	\includegraphics[width=2.7in]{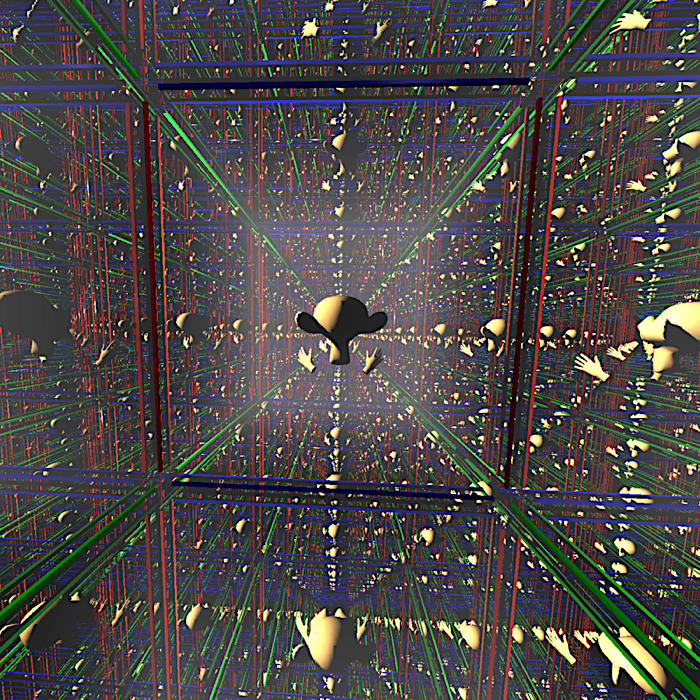}
	\vspace{-0.3cm}
	\caption{Inside view of the 3D flat torus using ray tracing~\cite{tr-09-19}.}
	\label{fig:torus_ray_tracing}
\end{figure}

Berger et al.~\cite{vc-rtorb-2014} was the first to proposed an \textit{image-based} algorithm  to visualize non-Euclidean spaces. Their rendering algorithm exploited programmable compute shaders and CUDA to implement ray-tracing on the GPU. However, the algorithm was limited to render only scenes composed of Lambertian surfaces described as implicit surfaces. Shadows and reflections were not simulated.   
In \cite{tr-09-19}, we proposed the first real-time visualization of such spaces using RTX GPUs. More complexes scenes were rendered, and shadows and reflections were introduced (Figure~\ref{fig:torus_ray_tracing}).

We take the challenge of computing the global illumination in non-Euclidean spaces using a path tracer. In such spaces, the concept of photorealistic rendering is not well defined because there is no real camera there. However, the possibility of defining a \textit{global illumination equation} in non-Euclidean geometry opens the door for a path tracing generalization. The images are inspiring.

\subsection{Non-Euclidean Spaces}\label{s-nonEuclidean}\hfill\\
In dealing with path tracing algorithms is required: (1) Space must be locally similar to a Euclidean space --- a \textit{manifold}. This allows us to model the camera and scene objects;  (2) For each point $p$ we need \textit{tangent} vectors pointing in all directions. The \textit{inner product} between two tangent vector is required to simulate light effects; (3) For a point $p$ and a vector $v$ tangent at $p$, we need a \textit{ray}, and its intersections with objects in the scene.

\textit{Geometric manifolds} satisfies the above properties. Such objects are locally geometrical similar to special spaces called \textit{model geometries}. In dimension two, for example, there are exactly three models: Euclidean, hyperbolic, and spherical spaces. In dimension three, there are five more model geometries, however, in this work, we focus on the (classical) first three spaces. We describe these topics in more detail below.
Great texts on this subject are Thurston~\cite{thurston1979} and Martelli~\cite{martelli2016introduction}. 

\begin{definition}[Euclidean space]
\label{subsection:euclian_space}
\normalfont
The \textit{Euclidean space} $\mathbb{E}^3$ is $\R^3$ endowed with the classical \textit{inner product} $\langle u,v \rangle_{\E}=u_xv_x+u_yv_y+u_zv_z$ where $u$ and $v$ are vectors in $\R^3$. 
The \textit{distance} between two points $p$ and $q$ is $d_\mathbb{E}(p,q)=\sqrt{\langle p-q,p-q \rangle_{\E}}$. 
The curve $\gamma(t)=p+tv$ describes a \textit{ray} leaving a point $p$ in a direction $v$.
Analogously, for any $n>0$ the Euclidean space $\mathbb{E}^n$ is constructed.
\end{definition}

\begin{definition}[Hyperbolic space]
\label{subsection:hyperbolic_space}
\normalfont
The \textit{Lorentzian inner product} of the vectors $v$ and $u$ in $\R^4$ is $\hDot{u}{v}=u_x v_x+u_y v_y+u_z v_z-u_w v_w$. The \textit{Lorentzian space} is $\R^4$ endowed with this product.
The \textit{hyperbolic space} $\mathbb{H}^3$ is the hyperboloid $\{p\in \R^4|\,\,\hDot{p}{p}=-1\}$ endowed with a special metric $d_{\H}(p,q)=\cosh^{-1}(-\hDot{p}{q})$.

A tangent vector $v$ to a point $p$ in $\mathbb{H}^3$ satisfies $\hDot{p}{v}=0$, then the \textit{tangent space} $T_p\H^3=\{v\in\R^4|\,\, \hDot{p}{v}=0\}$.
\textit{Rays} in $\mathbb{H}^3$ are the intersections between $\mathbb{H}^3$ and the planes in $\R^4$ containing the origin. For instance, the ray leaving a point $p\in\mathbb{H}^3$ in a tangent direction $v$ is the intersection between $\mathbb{H}^3$ and the plane spanned by the vectors $v$ and $p$ in $\mathbb{E}^4$. Such ray can be parameterized as $r(t)=\cosh(t)p+\sinh(t)v$.

$\mathbb{H}^3$ contains no straight line, thus its rays can not be straight. However, it is possible to model $\H^3$ in the unit open ball in $\R^3$ --- known as \textit{Klein model} $\mathbb{K}^3$--- such that the rays are straight lines. More precisely, each point $p\in\H^3$ is projected in the space $\{(x,y,z,w)\in\R^4| \,\, w=1\}$ by considering $p/p_w$. 
\end{definition}

\begin{definition}[Elliptic Space]
\label{subsection:sphere}
\normalfont
The \textit{$3$-sphere} $\S^3$ is the set $\{p\in \mathbb{E}^4|\,\,\eDot{p}{p}=1\}$ endowed with the metric $d_{\mathbb{S}}(p,q)=\cos^{-1}{\langle p,q \rangle}_\E$. A tangent vector $v$ to a point in $\S^3$ satisfies $\eDot{p}{v}=0$, then the \textit{tangent space} $T_p\S^3=\{v\in \E^4|\eDot{p}{v}=0\}$. The space $T_p\S^3$ inherits the Euclidean inner product of $\E^4$.
A \textit{ray} in $\mathbb{S}^3$ passing through a point $p$ in a tangent direction $v$ is the intersection between $\mathbb{S}^3$ and the plane spanned by $v$, $p$, and the origin. Such ray can be parameterized as $r(t)=\cos(t)p+\sin(t)v$.
\end{definition}

A \textit{$3$-manifold} $M$ is a topological space which is locally identical to the Euclidean space $\mathbb{E}^3$: there is a neighborhood of every point in $M$ mapped diffeomorphically to the open ball of $\mathbb{E}^n$. These maps are called \textit{charts}. The change of charts between two neighborhoods in $M$ must be differentiable. 
Examples of $3$-manifolds include the Euclidean, hyperbolic, and spherical spaces. 

\textit{Metric} and \textit{rays} are fundamental objects when working with path tracing, since light travels along with rays and are distributed in the scene surfaces by the metric. A manifold $M$ endowed with a metric at each tangent space (in a smooth manner) is called \textit{Riemannian}. This allows us to compute the length of \textit{vectors} and distances between points in $M$. A \textit{geodesic} (ray) in $M$ is a curve locally minimizing length.

We remember an algebraic (very computational) way to construct manifolds from simple ones.
Let $M$ be a connected manifold and $\Gamma$ be a discrete group of isometries acting on $M$, the \textit{quotient} $M/\Gamma$ is the set $\{\Gamma \cdot p|\,\, p \in M\}$ where $\Gamma\cdot p=\{g(p)|\,\, g\in\Gamma\}$ is the \textit{orbit} of $p$. For example, the quotient of $\mathbb{E}^3$ by the group of translation gives rise to the \textit{flat torus} $\mathbb{T}^3$.
A ray $r$ leaving a point $p\in \mathbb{T}^2$ in a direction $v$ is described by considering the fractional part of the coordinates of $r(t)=p+t\cdot v$.

Interesting cases arise when $M$ is a \textit{geometry} model: Euclidean, hyperbolic, and spherical. The space $M/\Gamma$ inherits the \textit{geometric structure} of $M$. For example, $\mathbb{T}^3$ has the geometric structure modeled by $\mathbb{E}^3$.
The \textit{fundamental domain} $\Delta$ plays an important role in the above construction being the region of $M$ containing exactly one point for each orbit. The unit cube is the fundamental domain of $\mathbb{T}^3$.


\section{Illumination of Non-Euclidean Spaces}\label{s-visualization_nonEuclidean}
This section focus on extending the local and global illumination equations to a Riemannian manifold $(M,g)$, the next section presents a path tracer algorithm for the class of geometric manifolds.
We consider that light propagates along with rays in $(M,g)$, and that it is constant in a vacuum.
Then inside views of a scene inside $M$ can be rendered by tracing rays: given a point (eye) and a direction (pixel), we trace a geodesic (first ray). When it reaches a point on a surface object, we compute its \textit{Riemannian illumination} by considering the \textit{direct} and \textit{indirect} light contributions. The direct illumination considers the rays coming directly from light sources and the indirect gathers the light rays bounced by other surfaces.

\textit{Illumination} is the process of simulating the light leaving a point in a given direction. Classic approaches to perform such tasks are not suited for Riemannian manifolds, so we propose a more general illumination definition.

Let $\mathcal{C}$ be the space of RGB colors. The \textit{Riemannian illumination} of a $3$-manifold $M$, with a scene embedded on it, is a function $L:M\times\S^2\to \mathcal{C}$, where for each point $p\in M$ and unit tangent direction $v\in T_pM$ we have a color $L(p,v)$ representing the light leaving this direction. 
Because we are considering the light constant along geodesics, the function $L$ only needs to be computed on the scene surface objects.

There are many approaches to compute the illumination function. We start with a local model considering only the direct contributions of the light sources and surfaces composed of diffuse and specular materials. Then we incorporate the indirect light contribution generating a global model.

\subsection{Riemannian local illumination}\hfill\\
We start with the classical local illumination in a Euclidean scene. Let $p$ be a point in a surface object $S$, and $N$ its normal vector. Let $l$ be a point light, $w_i=(l-p)/\eNorm{l-p}$ is the \textit{incident vector} and $w_r = -w_i + 2\eDot{w_i}{N}N$ is the \textit{perfect reflection} of $w_i$. For a vector $v$ leaving $p$, the \textit{local illumination function} is:
\begin{equation}\label{eq:phong}
    L(p,v) = k_aL_a + k_d\sum_{j=1}^{n}L_i\eDot{w_i^j}{N} + k_s\sum_{j=1}^{n}L_i\eDot{w_r^j}{v}^{k_e},
\end{equation}
where $L_a$ indicates the ambient light intensity, $L_i$ is the radiance emitted by the light source $l_i$, and the numbers $k_a$, $k_d$, $k_s$, and $k_e$ represent the \textit{ambient}, \textit{diffuse}, \textit{specular}, and \textit{roughness} coefficients of the surface $S$ at $p$.

We now extend Equation~\ref{eq:phong} to a Riemannian manifold $(M,g)$; $g$ is the Riemannian metric.
Here $p$ is a point in a smooth surface $S$ embedded in $M$, and $N$ is the normal vector at $p$. The direction $w_i$ is the unit tangent vector at time zero of the geodesic running from $p$ towards the light source $l$. Then $w_r = -w_i + 2g_p(w_i,N)N$ is the perfect reflection of $w_i$.
The metric $\eDot{}{}$ used in Equation~\ref{eq:phong} is replaced by the Riemannian metric $g_p{}{}$ of $M$ restricted to $p$, this gives rise to the \textit{Riemannian local illumination} for a unit direction $v$ at $T_pM$:
\begin{equation}\label{eq:phong_Rie}
    L(p,v) = k_aL_a + k_d\sum_{j=1}^{n}L_ig_p(w_i^j,N) + k_s\sum_{j=1}^{n}L_ig_p(w_r^j,v)^{k_e},
\end{equation}

To render an image of a scene embedded in the Riemannian manifold $M$ we define a \textit{Riemannian  ray casting}: it shoots rays through the pixels, if a ray intersects a surface, we compute its color using Riemannian local illumination function.
Consider a $2$-sphere $\S^2_o$ centered in a point $o$ (the observer) in $M$. We give a color for each sphere point (ray direction) in the observer field of view $V$ by tracing a ray; In other words, $V \subset \S^2_o$ carries the image. We call this procedure \textit{Riemannian ray casting}.
Specifically, the unit sphere $\S^2_o$ is centered at the origin of $T_oM$. For each direction $v$ in $\S^2_o$ we attribute a color $c$ by launching a ray $\gamma$ from $o$ in the direction $v$. Each time $\gamma$ intersects a scene object at a point $p=\gamma(t)$ we compute its Riemannian local illumination $L(p, -\gamma'(t))$.

Before we go to a global illumination function we observe in Equation~\ref{eq:phong_Rie} that the ambient term $k_aL_a$ is a very rough approximation of the indirect light contribution. The other terms in Equation~\ref{eq:phong_Rie} is the direct light contribution when dealing with diffuse/specular surfaces. 

\subsection{Global illumination in non-Euclidean manifolds}\hfill\\
We remember the classical \textit{global illumination function} of the Euclidean space $\E^3$ introduced by Kajiya~\cite{kajiya1986rendering}. For each point $p$, it computes the amount of light emitted in a direction $v$. It is modeled through the integral equation over the hemisphere $\Omega(p)=\{v\in\S^2|\, \eDot{v}{N(p)\}\geq 0}$:
\begin{equation}\label{eq:global}
    L(p,v) =L_e(p,v)+\int_{\Omega} f_r(p,v,w_i)L(p,w_i)\eDot{w_i}{N}dw_i.
\end{equation}
The \textit{bidirectional reflectance distribution function} (BRDF) $f_r(p,v,w_i)$ defines how the light reflects at $p$, and $L_e(p,w)$ is the light emitted at $p$ in the direction $v$. 

Equation~\ref{eq:global} generalizes Equation~\ref{eq:phong} because it integrates over all direction entering the hemisphere $\Omega$ while Equation~\ref{eq:phong} considers only the directions coming directly from the light points. The other indirect directions are roughly approximated by the ambient term $k_aL_a$. 

Again, replacing the Euclidean metric with the Riemannian metric, we obtain a global illumination function for a Riemannian manifold $(M,g)$,
\begin{equation}\label{eq:global_Riem}
    L(p,v) =L_e(p,v) + \int_{\Omega} f_r(p,v,w_i)L(p,w_i)g_p(w_i,N)dw_i.
\end{equation}
Taking into account the import contribution of the directions $w_i$ leaving the point $p$ in the direction of the light points $l_i$, we divide Equation~\ref{eq:global_Riem} into the direct and indirect components $L(p,v) =L_e(p,v) + L_{dir}(p,v) +L_{ind}(p,v)$. The direct contribution $L_{dir}(p,v)$ is computed using Riemannian local illumination function (Equation~\ref{eq:phong_Rie}). For the indirect contribution $L_{ind}(p,v)$, we use \textit{Monte Carlo integration} to estimate a good approximation.

Let $\{w_i^1,\ldots,w_i^n\}$ be $n$ vectors on the hemisphere $\Omega$ chosen using a distribution density $d_w$, the Monte Carlo integration states that
\begin{equation}\label{eq:global_Riem_Approx}
    L_{ind}(p,v) \approx \frac{1}{n}\sum_{k=1}^n \frac{f_r(p,v,w_i^k)L(p,w_i^k)g_p(w_i^k,N)}{d_w(w_i^k)}.
\end{equation}
If $L(p,w_i^k)$ is known, the \textit{law of large number} ensures the convergence when $n\to\infty$. Otherwise, computing these approximation for each $L(p,w_i^k)$, and repeating such procedure for a finite number $m$ of times, we obtain a version of the \textit{path tracing} algorithm. 
It is well-known in compute graphics, since Kajiya~\cite{kajiya1986rendering}, that the number $m$ does not need to be large for computing realistic images. 


\section{Path tracing in non-Euclidean spaces}\label{s-path_tracer}
We combine Riemannian ray casting and Riemannian global illumination to synthesize inside views of non-Euclidean spaces. 
This generalizes the classical \textit{path tracing}. We focus on a geometric manifold $M/\Gamma$ since it admits a well-behaved geometry $M$ and a combinatorial description of its topology in terms of $\Gamma$.

Our method approximates the Riemannian global illumination of the visible surfaces inside a geometric manifold using the ray tracing capabilities of the RTX platform.
We will discuss first the basic principles of path tracing in Non-Euclidean spaces, as well as, the general algorithm in CPU. We will show how to map the computation to the RTX pipeline and present the details of GPU implementation.





\subsection{Overview of the Method}\hfill\\
The path tracing is the most natural method to produce photorealistic images of Euclidean scenes. Thus extending the technique to non-Euclidean spaces would provide a flavor of the photo-realism inside such abstract spaces. It is necessary to take into account the geometry/topology of the non-Euclidean space. The first aspect of this task is to simulate the ray path as it travels inside the space, starting from the point of observation until it intersects with a visible object --- the Riemannian ray casting. The second aspect amounts to the illumination, evaluating the light scattered from the environment in the ray direction --- the Riemannian global illumination. Because of the non-trivial topology, the ray path is updated as it exits the fundamental domain.

\subsection{Algorithm in CPU}\hfill\\
We present the basic path tracing algorithm for a geometric manifold $M/\Gamma$, and compare it with the traditional path tracing of Euclidean space. 

We start with an algorithm to trace rays inside $M/\Gamma$. 
Let $p$ be a point inside the fundamental domain $\Delta$ of $M/\Gamma$, and $v$ be a direction at $T_pM$, Algorithm~\ref{alg:trace-ray} trace a ray $\gamma$ from $p$ towards the direction $v$. If $\gamma$ intersects a scene object in  $\Delta$, the point and direction are updated. 
\begin{algorithm}[!h]
\KwData{point $p$, direction $v$}
\KwResult{bool $hit$ }
\SetAlgoLined
Trace a ray $\gamma$ from $(p, v)$ inside $\Delta$\;
\Repeat{$i\le maxlevel$}{
Find closest intersection $\gamma(t)$ with objects $O$ in $\Delta$\;
\eIf{$\gamma(t) \ne \emptyset$}{  
     Update $p=\gamma(t)$ and $v=\gamma'(t)$\;
     \Return true\;
}{
     Find intersection of $\gamma$ with faces of $\Delta$ \;
     Compute the new origin $p'$ and ray $\gamma'$\;
     $++i$;
 }
 }
 \Return false\;
\caption{Trace ray}
\label{alg:trace-ray}
\end{algorithm}

As it can be verified in Algorithm~\ref{alg:trace-ray}, the rays are intersected with visible objects (line 5) and if there is a hit (line 6), the ray point and direction is updated (line 7).
The whole computation has the {\em fundamental domain} as a base, which is modeled by a polyhedron $\Delta$. Therefore, as the ray hits a face $F$ of $\Delta$ (line 9), we transport it using the corresponding transformation of the discrete group $\Gamma$ (line 10--11). 
This is the most important and critical step since it
depends on the geometry and topology of the space. As such, it is specific for each type of space.
For practical computational reasons, we cannot continue the ray indefinitely, thus a maximum level is set to stop the path (line 15). 

Algorithm~\ref{alg:path_tracing} describes the path tracer. The rays are generated from the observer's point of view (lines 1--2). Using Algorithm~\ref{alg:trace-ray} the visible points are computed and if there is a hit (line 3), shading is done using direct and indirect illumination (line 4--5). Note that Algorithm~\ref{alg:path_tracing} is computing the Riemannian ray casting. To compute a good approximation of the image, it is common to accumulate the resulting image of Algorithm~\ref{alg:path_tracing} and line 2 chooses the direction associated with the pixel using a distribution density function.
\begin{algorithm}[!h]
\SetAlgoLined
\For {each pixel $\sigma \in I$}{
    Let $p:=0$ and $v$ be the direction associated to $\sigma$\;
    \If{TraceRay$(p, v, d)$}
    {
        Define a depth $d$ and a color $c=0$\;
        Shade $\sigma$ using $L_{dir}(p, v)+L_{ind}(p, v, d, c)$\;
    }
}
\caption{Path Tracing in non-Euclidean spaces}
\label{alg:path_tracing}
\end{algorithm}

The direct illumination is computed using Equation~\ref{eq:phong_Rie}. To add the shadows we multiply the direct illumination by a factor $0$ if tracing a ray from the hit point to the light source a scene intersection occurs, and $1$ otherwise.

The indirect illumination is presented in Algorithm~\ref{alg:indirect-illumination}. An integer variable \textit{depth} $d$ is considered to control the number of ray light bounces of the path tracer. If $d$ is greater than zero and the ray launched from $(p,v)$ intersect the scene (line 3), we sample a hemisphere direction and compute an approximation of the Riemannian global illumination (line 4--5). Clearly, in line 5 Algorithm~\ref{alg:indirect-illumination}, is called again. This came from Equation~\ref{eq:global_Riem_Approx} considering only one sample.
\begin{algorithm}[!h]
\KwData{point $p$, direction $v$, depth $d$, color $c$}
\KwResult{color $c$ }
\SetAlgoLined
\If{ $d\geq0$ and TraceRay$(p, v)$}
{  
    Sample a direction $w\in\Omega(p)$ using a distribution $d_w$\;\vspace{0.1cm}
    c+=$L_{dir}(p, v)+\displaystyle\frac{f_r(p,v,w)L_{ind}(p,w, d-1, c)g_p(w,N)}{d_w(w)}$\;\vspace{0.1cm}
}
\Return c\;
\caption{Indirect illumination}
\label{alg:indirect-illumination}
\end{algorithm}


\subsection{RTX Pipeline}\hfill\\
NVidia RTX is a hardware/software platform with support for real time ray tracing. The ray tracing code of an application using this architecture consists of CPU host code, GPU device code, and the memory to transfer data to the Acceleration Structures for fast geometry culling when intersecting rays with scene objects.
Specifically, the CPU host code manages the memory flow between devices, sets up, controls and spawn GPU shaders and defines the Acceleration Structures. 

These shaders correspond to tasks in Algorithm~\ref{alg:trace-ray}, \ref{alg:path_tracing}, and \ref{alg:indirect-illumination}. The {\em Ray Generation Shader} is responsible for creating the rays (line 1 in Algorithm~\ref{alg:path_tracing}), which are defined by their origins and directions (line 2). A call to TraceRay() launches a ray (line 3 in Algorithm~\ref{alg:trace-ray}). The next stage is a fixed traversal of the Acceleration Structure described only at a high level here. 
This traversal uses an {\em Intersection Shader} to calculate the intersections (line 5  in Algorithm~\ref{alg:trace-ray}). All hits found pass by tests to verify if they are the closest hit. 
After no additional hits are found, the {\em Closest-Hit Shader} is called for the closest intersection point (line 7 in Algorithm~\ref{alg:trace-ray}). 
In case no hits are found, Miss Shader is called as a fallback case. It is important to note that additional rays can be launched in the Closest-Hit and Miss shaders. 

The {\em scene objects} and the {\em boundary of the fundamental domain} are treated differently when mapping the algorithm to the RTX pipeline --- while the scene objects are tested and shaded in the regular way (lines 5 and 7  in Algorithm~\ref{alg:trace-ray}), the boundary of the fundamental domain is used to transport the rays by the discrete group (lines 9 and 10). This is implemented with a custom designed Miss Shader.

\subsection{GPU Implementation}\hfill\\
\label{subsection:implementation}
The implementation of our visualization platform in GPU is built on top of Falcor using DirectX 12 on Windows 10.
The Falcor development framework consists of a library with support for DXR at a high level and a built-in scene description format. 
We use the software Blender to create the scene objects.

The core functionality of our system's architecture consists of a set of shaders that are mapped to the RTX GPU pipeline as described above. We developed generic shaders for each stage of the GPU path tracing pipeline that are independent of the geometric structure of the non-Euclidean space.

{\bf Ray Generation Shader}: Creates camera rays using the isometries of the space to transform the ray origin and direction to the camera coordinate system.

{\bf Intersection Shader}: Computes the ray-object intersection using a parameterization of the ray. Ray and objects are defined based on the model Geometry.

{\bf Closest Hit Shader}: Performs the shading operation. This includes computing the local and global illumination. The local illumination amounts to the direct contribution of light sources that are based on angles between the light direction and the surface normal, as well as, the distance to the light. The global illumination is computed by launching indirect rays based on the surface BRDF.

{\bf Miss Shader}: Deals with the transport of rays in the covering space, as they leave the fundamental domain. For this, the rays are tested for intersection with the boundary of the polyhedron $\Delta$.



\newpage
\section{Examples and Results}\label{s-examples}
We present some expressive output images from our
implementation of the algorithm in GPU using RTX. We focus on visualizing the 3D flat torus.

\subsection*{Flat Torus}\hfill\\
\label{subsection:flat-torus}
Probably the most famous and easiest example of a compact $3$-manifold is the \textit{flat} \textit{torus} $\T^3$. Topologically, it is obtained by gluing opposite faces of the unit cube $[0,1]\times[0,1]\times[0,1] \subset \mathbb{E}^3$. It is easy to check that the neighborhood of each point in $\mathbb{T}^3$ is a $3$-ball of the Euclidean space. Thus $\mathbb{T}^3$ is indeed a $3$-manifold.

$\T^3$ is modeled by $\E^3$ because it is the quotient of the Euclidean space by the group of translation spanned by $(x,y,z)\to(x\pm 1,y,z)$, $(x,y,z)\to(x,y\pm 1,z)$, and $(x,y,z)\to(x,y,z\pm 1)$. Thus, the unit cube is the fundamental domain of $\T^3$. 

\begin{figure}[!h]
\centering
\includegraphics[width=4.9in]{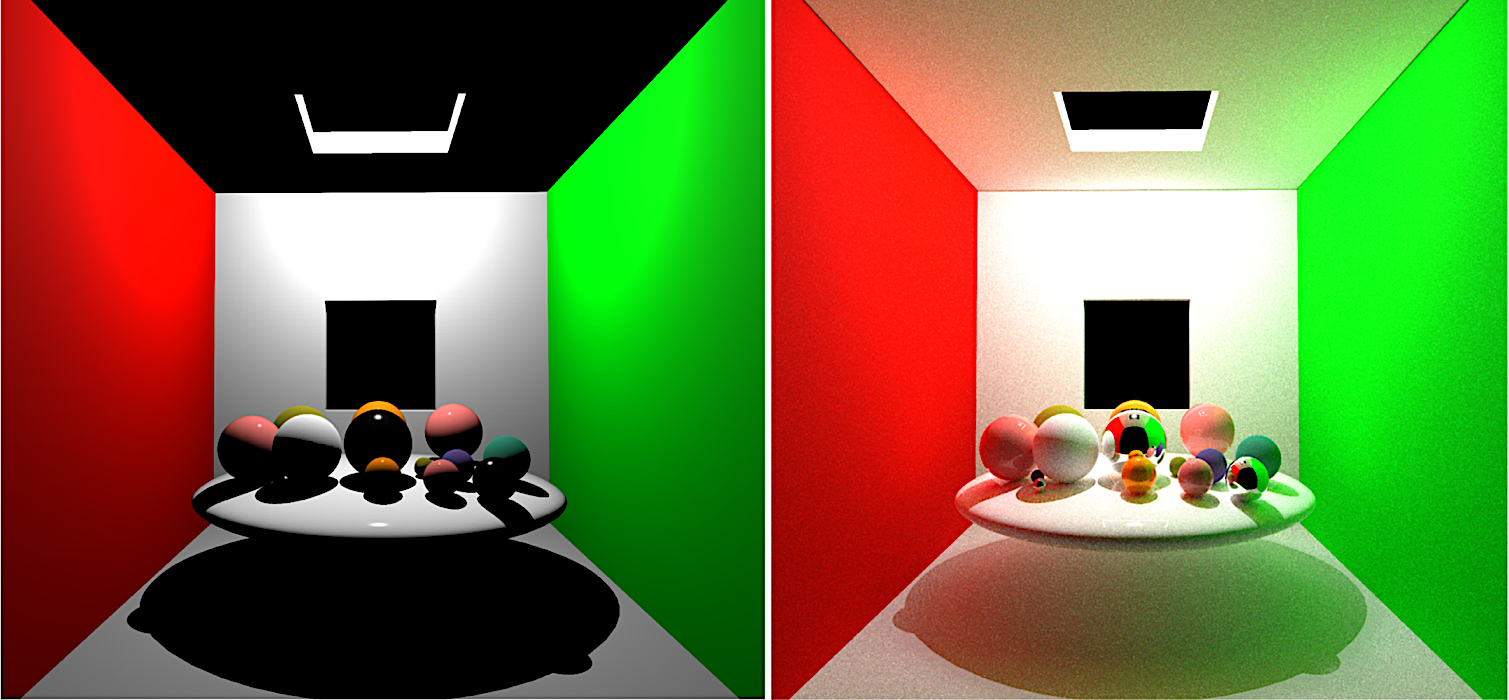}
\caption{\small Cornell box inside the fundamental domain. Rendered considering zero/five bounces.}
\label{fig:fundamental_domain}
\end{figure}

Figure~\ref{fig:fundamental_domain} presents the visualization of a scene inside the fundamental domain using the path tracing algorithm. On its left, we do not use the indirect light contribution, on the right we add the indirect light contribution considering five bounces. Note that we are using a \textit{cornel box} with a additional window to set up our scene. The spheres inside the cube have specular, diffuse, and composed materials. The point light is above the box, close to the top window.

We now identify the boundaries of the torus fundamental domain (the cube).
A ray leaving a point $p\in \mathbb{T}^3$ in a direction $v$ is parameterized as $r(t)=p+t\cdot v$ in $\mathbb{E}^3$. For each intersection between $r$ and a face $F$ of the unit cube, we update $p$ by its correspondent point $p-n$ in the opposite face, where $n$ is the unit vector normal to $F$. The ray direction $v$ does not need to be updated.
The rays in $\T^3$ can return to the starting point, providing many copies of the scene. The immersive perception is $\E^3$ tessellated by unit cubes: each cube contains one copy of the scene.

\begin{figure}[!h]
\centering
	\begin{tabular}{c}
	    \includegraphics[width=4.9in]{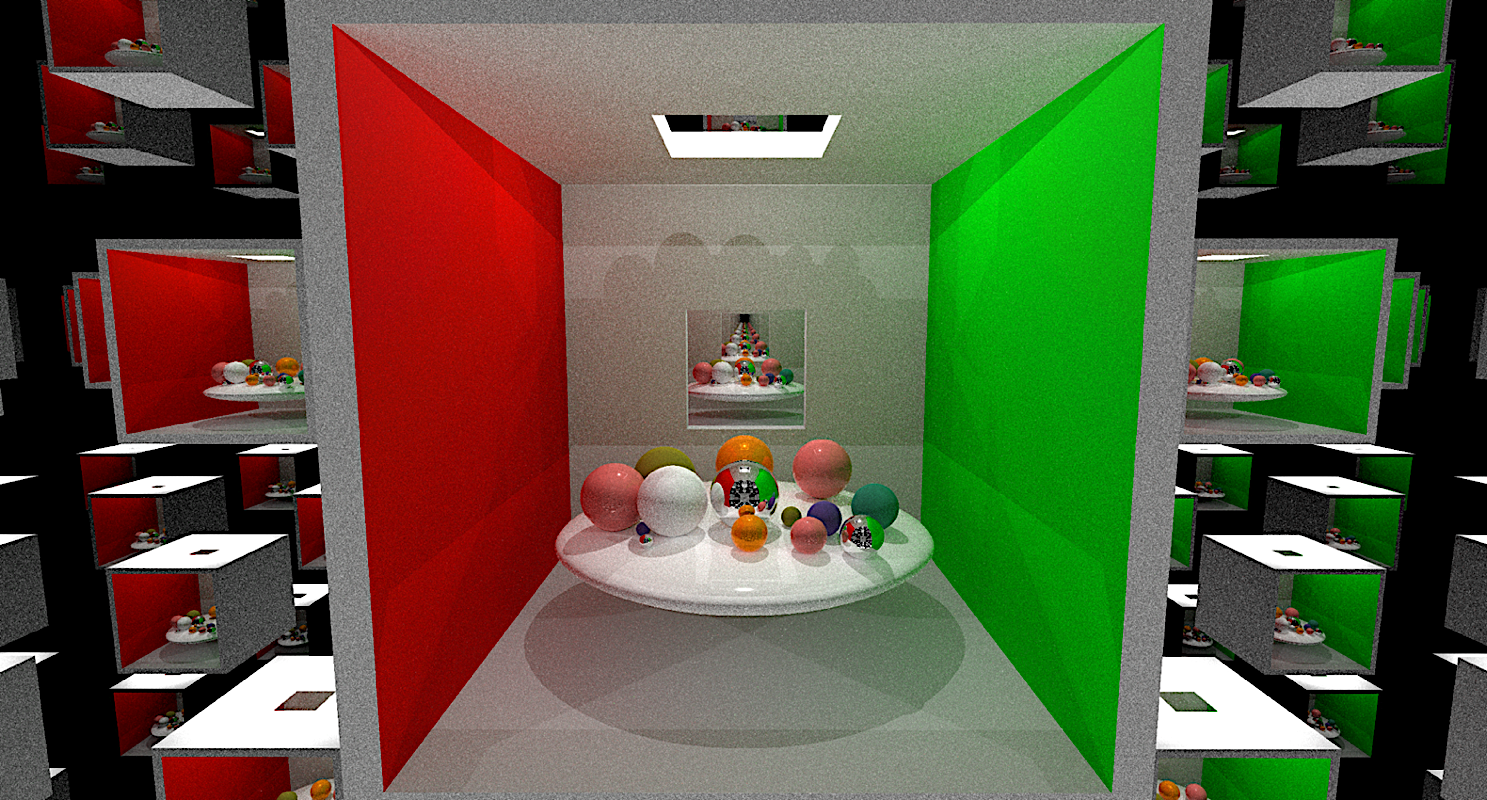} \\
	    \includegraphics[width=4.9in]{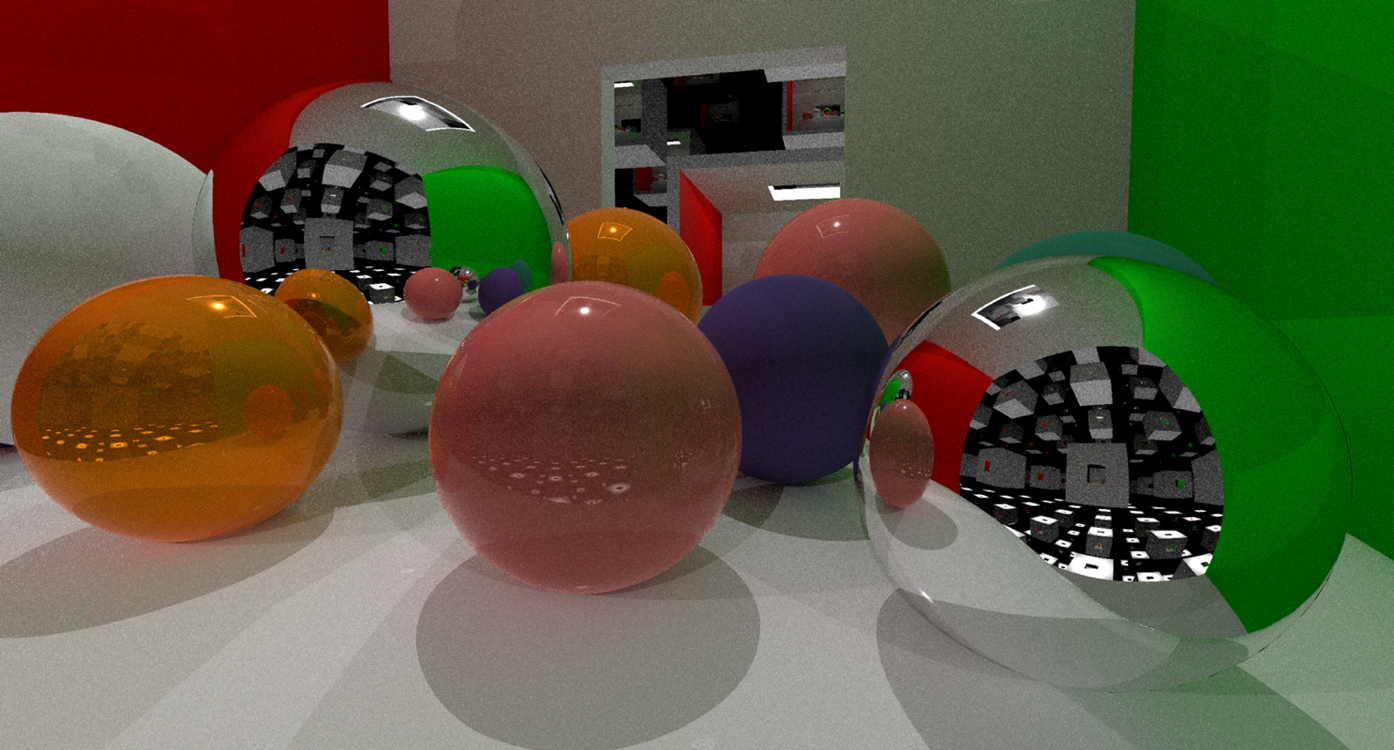}
	\end{tabular}
\caption{\small Immersive view in the 3D flat torus. The face pairing makes the rays that leave a face return from its opposite face, giving rise to many copies of the scene.}
\label{f-3torus}

\end{figure}

Figure~\ref{f-3torus} provides an immersive visualization of the $3$-dimensional torus $\mathbb{T}^3$ using the path tracer algorithm. On its top, an outside view of the embedded cornel box showing how the fundamental domain tesselates the space. On its bottom, a closer view on the specular spheres embedded in the scene to show the tesselation being reflected on their surface.

\subsection*{Mirrored Dodecahedron}\label{subsection:m-dodeca}\hfill\\
For an example of a space modeled by the hyperbolic space, consider the dodecahedron embedded in $\mathbb{H}^3$. Let $\Gamma$ be the group of reflections generated by the dodecahedral faces. With an appropriate scale, the dihedral angle of the dodecahedron reaches $90$ degrees. The quotient $\mathbb{H}^3/\Gamma$ is the \textit{mirrored dodecahedral space}. $\Gamma$ tessellates $\H^3$ with dodecahedra, each edge has exactly $4$ cells. 

\begin{figure}[h]
\centering
\includegraphics[width=4.9in]{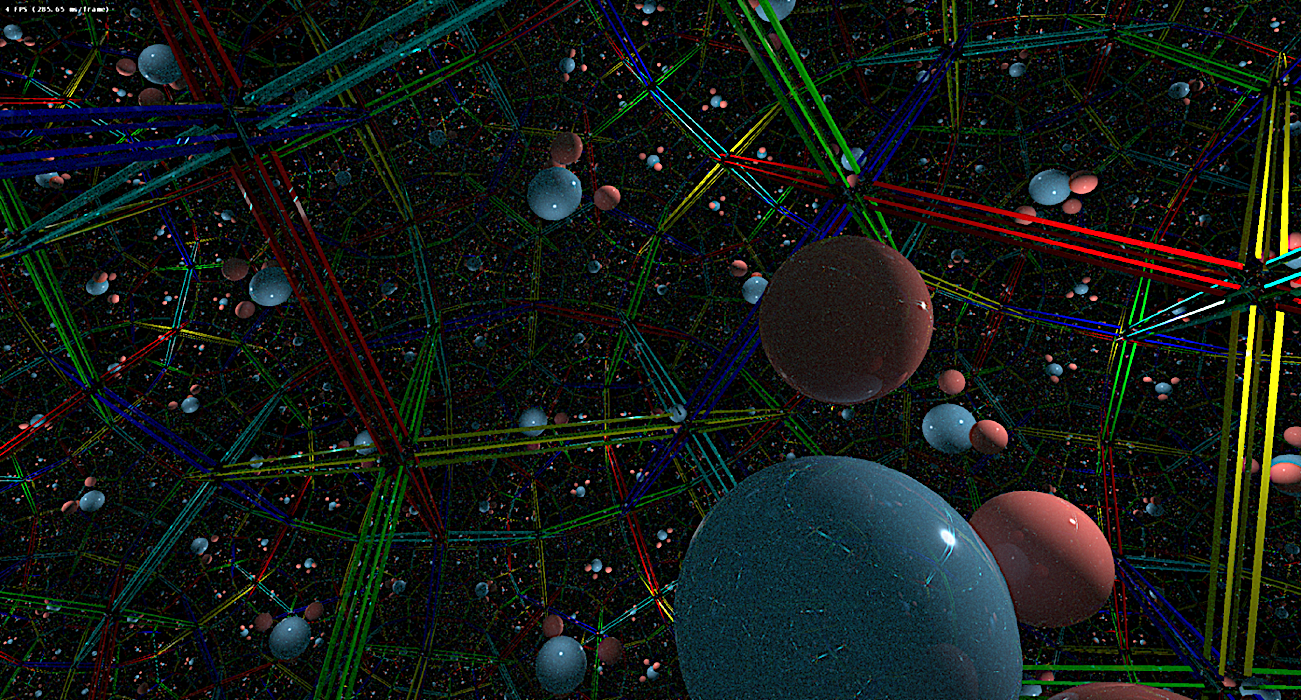}
\caption{\small Immersive visualization of the mirrored dodecahedron. This space is obtained by considering the faces of a hyperbolic regular dodecahedron to be perfect hyperbolic mirrors.}
\label{f-mirror-dodeca}
\end{figure}

Figure~\ref{f-mirror-dodeca} illustrates an immersive visualization of the mirrored dodecahedron using the reflection definition in the hyperbolic space. There are exactly three spheres, two red and a blue. We add the dodecahedron edges to highlight the hyperbolic space tessellation. The image is the inside view of the group of reflection acting on the Hyperbolic space.

\bibliographystyle{amsplain}
\bibliography{ray3d}

\providecommand{\bysame}{\leavevmode\hbox to3em{\hrulefill}\thinspace}
\providecommand{\MR}{\relax\ifhmode\unskip\space\fi MR }
\providecommand{\MRhref}[2]{%
  \href{http://www.ams.org/mathscinet-getitem?mr=#1}{#2}
}
\providecommand{\href}[2]{#2}
\begin{thebibliography}{1}

\bibitem{vc-rtorb-2014}
Pierre Berger, Alex Laier, and Luiz Velho, \emph{An image-space algorithm for
  immersive views in 3-manifolds and orbifolds}, Visual Computer (2014).

\bibitem{kajiya1986rendering}
James~T Kajiya, \emph{The rendering equation}, Proceedings of the 13th annual
  conference on Computer graphics and interactive techniques, 1986,
  pp.~143--150.

\bibitem{martelli2016introduction}
Bruno Martelli, \emph{An introduction to geometric topology}, arXiv preprint
  arXiv:1610.02592 (2016).

\bibitem{pharr2016physically}
Matt Pharr, Wenzel Jakob, and Greg Humphreys, \emph{Physically based rendering:
  From theory to implementation}, Morgan Kaufmann, 2016.

\bibitem{thurston1979}
W.P. Thurston, \emph{The geometry and topology of three-manifolds}, Princeton
  University, 1979.

\bibitem{tr-09-19}
Luiz Velho, Tiago Novello, Vinicius Silva, and Djalma Lucio,
  \emph{Visualization of non-euclidean spaces using ray tracing}, Technical
  Report TR-09-2019, VISGRAF Lab - IMPA, 2019.

\end{thebibliography}

\end{document}